\documentstyle[prl,aps,epsf]{revtex}
\draft
\begin{document}
\twocolumn[\hsize\textwidth\columnwidth\hsize\csname @twocolumnfalse\endcsname

\title{Metastability and Transient Effects in Vortex Matter
Near a Disorder Driven Transition}
\author{C.~J. Olson, C. Reichhardt, R.~T.~Scalettar, G.~T.~Zim{\' a}nyi}
\address{Department of Physics, University of California, Davis, California
95616.}

\author{Niels Gr{\o}nbech-Jensen}
\address{Department of Applied Science, University of California,
Davis, California 95616.\\
NERSC, Lawrence Berkeley National Laboratory, Berkeley, California 94720.}

\date{\today}
\maketitle
\begin{abstract} 
We examine metastable  
and transient effects both above and below the first-order disorder
driven decoupling line in a 3D simulation of magnetically interacting
pancake vortices. We observe pronounced transient and history effects  
as well as 
supercooling and superheating between the ordered and disordered phases.
In the disordered supercooled state 
as a function of DC driving, reordering occurs through the
formation of growing moving channels of the ordered phase. 
We find
that hysteresis in $V(I)$ is strongly dependent on the  
proximity to the decoupling transition line.
\end{abstract} 
\vspace{-0.1in}
\vspace{-0.4in}

\vskip2pc]
\narrowtext

Vortices in superconductors represent an ideal system in which to study
the effect of quenched disorder on elastic media.
The competition between 
the flux-line interactions, which order the vortex lattice, and the defects in
the sample, which disorder the vortex lattice, produce a remarkable 
variety of collective behavior \cite{Review1}. 
One prominent example is the peak effect in low temperature 
superconductors, which appears near $H_{c2}$ when a
transition from an ordered to a strongly pinned disordered state
occurs in the vortex lattice 
\cite{Kes2,Shobo3,Chaddah4,Henderson5,Andrei6,Xiao7,Paltiel8}.  
In high temperature superconductors, particularly BSCCO samples, a striking 
``second peak'' phenomenon is observed  
in which a dramatic increase in the critical current occurs 
for increasing fields.  It has been proposed that this is an
order-disorder or 3D to 2D transition.  
\cite{Giamarchi9,Cubitt10,Tamegai11,Glazman12,Feigelman13}

Recently there has been renewed interest in
transient effects, which have been observed in voltage response
versus time curves in low temperature superconductors 
\cite{Henderson5,Andrei6,Xiao7,Frindt14}. 
In these experiments the voltage response 
increases or decays with time, depending on how the vortex lattice
was prepared. The existence of
transient states suggests that the disordered phase can be {\it supercooled} 
into the ordered region, producing an increasing voltage response, whereas
the ordered phase may be {\it superheated} into the disordered region, giving
a decaying response. 
In addition to transient effects,
pronounced memory effects and hysteretic V(I) curves have been
observed near the peak effect in low temperature materials.
\cite{Kes2,Chaddah4,Henderson5,Andrei6,Xiao7}.
Xiao {\it et al.} \cite{Xiao7} 
have shown that transient behavior can lead
to a strong dependence of the critical current on the current
ramp rate. 
Recent neutron scattering experiments 
in conjunction with ac shaking have provided more direct 
evidence of supercooling and superheating near the peak effect 
\cite{Ling15}. Experiments on BSCCO have revealed that the high field 
disordered state can be supercooled to fields well below the
second peak line \cite{Konczykowski16}. 
Furthermore, transport experiments in BSCCO have shown
metastability in the zero-field-cooled state near the second peak as well as 
hysteretic 
V(I) curves \cite{Portier17} and transient effects \cite{Giller18}.
Hysteretic and memory effects have also been
observed near the second peak in YBCO 
\cite{Kokkaliaris19,Bekeris20,Esquinazi21}.   

The presence of metastable states and superheating/supercooling effects
strongly suggests
that the order-disorder transitions in these different 
materials
are {\it first order} in nature. 
The many similarities
also point to a universal behavior between the peak effect
of low temperature superconductors and the peak effect and
second peak effect of high temperature superconductors. 

A key question in all these systems is the nature of the
{\it microscopic dynamics} of the vortices 
in the transient states; particularly, whether plasticity or 
the opening of flowing channels are involved \cite{Andrei6}. 
The recent experiments have made it clear that a proper 
characterization of the static and dynamic phase diagrams must
take into account these metastable states, and therefore an understanding
of these effects at a microscopic level is crucial.  
Despite the growing amount of experimental work on  
metastability and transient effects in vortex matter,
these effects have not yet been 
investigated numerically.  

In this work we present the first numerical study of  
metastability and transient effects in vortex matter near 
a disorder driven transition. 
We demonstrate that the simulations reproduce many experimental
observations, including superheating and supercooling effects, and
then link these to the underlying microscopic vortex behavior.
We consider magnetically interacting
pancake vortices driven through quenched point disorder. As a
function of interlayer coupling or applied field the model exhibits a sharp 3D
(ordered phase) to 2D (disordered phase) 
disorder-driven transition \cite{FirstPaper22}. 
By supercooling or 
superheating the ordered and disordered phases, 
we find increasing or decreasing
transient voltage response curves,
depending on the amplitude of the drive pulse and the proximity to the 
disordering transition. 
In the supercooled transient states a
growing ordered channel of flowing vortices forms.
No channels form in the
superheated region but instead the ordered state is homogeneously destroyed.
We observe memory effects when a sequence of pulses is applied,
as well as ramp rate dependence and hysteresis in the V(I) curves. 
The critical current we obtain depends on how the system is prepared.

We consider a 3D layered superconductor containing an equal number of
pancake vortices in each layer, interacting magnetically.
We neglect the Josephson coupling which is
a reasonable approximation for highly anisotropic materials.
The overdamped equation of motion for vortex $i$ at $T=0$ is 
$ {\bf f}_{i} = -\sum_{j=1}^{N_{v}}\nabla_{i} {\bf U}(\rho_{ij},z_{ij})
+ {\bf f}_{i}^{vp} + {\bf f}_{d} = {\bf v}_{i}$. 
The total number of 
pancakes is $N_{v}$, and $\rho_{ij}$ and $z_{ij}$ are the distance between
vortex $i$ and vortex $j$ in cylindrical coordinates. We impose 
periodic boundary conditions in the $x$ and $y$ directions and open
boundaries in the $z$ direction. The magnetic interaction energy
between pancakes is \cite{Clem22a,Brandt23}
\begin{eqnarray}
{\bf U}(\rho_{ij},0)=2d\epsilon_{0}
\left((1-\frac{d}{2\lambda})\ln{\frac{R}{\rho}}
+\frac{d}{2\lambda}
E_{1}\right) \ ,
\nonumber
\end{eqnarray}
\begin{eqnarray}
{\bf U}(\rho_{ij},z)=-s_{m}\frac{d^{2}\epsilon_{0}}{\lambda}
\left(\exp(-z/\lambda)\ln\frac{R}{\rho}+
E_{2}\right) \ , \nonumber
\end{eqnarray}
where $R = 22.6\lambda$, the maximum in-plane distance, 
$E_{1} = 
\int^{\infty}_{\rho} d\rho^{\prime} \exp(\rho^{\prime}/\lambda)/\rho^{\prime}$,
$E_{2} = 
\int^{\infty}_{\rho} d\rho^{\prime} \exp(\sqrt{z^{2}+\rho^{\prime 2}}/\lambda)/\rho^{\prime}$,
$\epsilon_{0} = \Phi_{0}^{2}/(4\pi\xi)^{2}$, $d=0.005\lambda$ is the
interlayer spacing,
$\lambda$ is the London penetration depth and $\xi=0.01\lambda$ is the 
coherence length. 
When the magnetic field $H$ increases, the distance $\rho$ between pancakes
in the same plane decreases, but the distance $d$ between planes is unchanged.
Thus we model $H$ by scaling the strength of the in-
and inter-plane interactions via the prefactor $s_{m}$,
such that $s_m \propto 1/{H}$.
We denote the coupling strength at which the sharp 3D-2D transition occurs
as $s_{m}^{c}$. 
We model the pinning
as $N_p$ short range attractive 
parabolic traps that are randomly distributed in each 
layer. 
The pinning interaction is 
${\bf f}_{i}^{vp} = \sum_{k=1}^{N_{p}}(f_{p}/\xi_{p})
({\bf r}_{i} - {\bf r}_{k}^{(p)})\Theta( 
(\xi_{p} - |{\bf r}_{i} - {\bf r}_{k}^{(p)} |)/\lambda)$,
where the pin radius 
$\xi_{p}=0.2\lambda$, the pinning force  
is $f_{p}=0.02f_{0}^{*}$, and $f_{0}^{*}=\epsilon_{0}/\lambda$.
Throughout this work we will use 16 layers in a
$16\lambda \times 16\lambda$ system with a vortex density of 
$n_v = 0.35/\lambda^2$ and a pin density of $n_p = 1.0/\lambda^{2}$ 
in each of the layers. There are 80 vortices per layer, giving 
a total of 1280 pancake vortices.

\begin{figure}
\center{
\epsfxsize = 3.2 in
\epsfbox{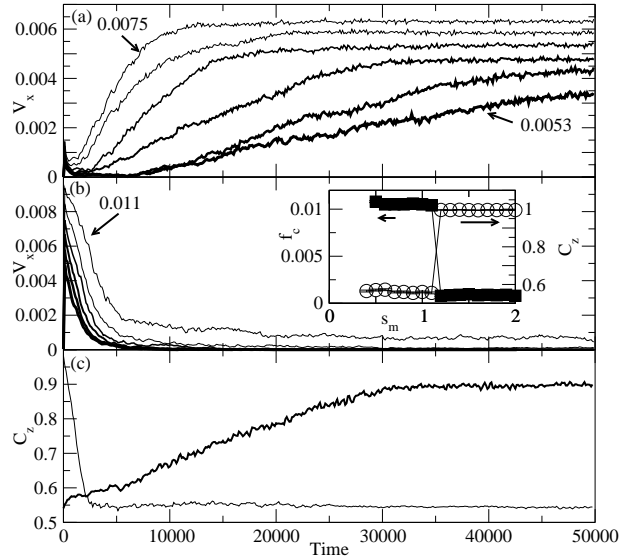}}
\caption{Transient voltage response curves $V_x$ versus time.
(a) Increasing response for
supercooled state, for $s_m= 2.0$ and $f_d/f_{0}^{*} =$ 
(bottom) 0.0053, 0.0055, 0.006,
0.0065, 0.007, and 0.0075 (top).
(b) Decreasing response for superheated state, for $s_m=0.7$ and 
$f_d/f_{0}^{*} =$ 
(bottom) 0.006, 0.007, 0.008, 0.009, 0.010, and 0.011 (top).
Inset to (b): Critical current $f_c$ (filled squares) and 
interlayer correlation $C_z$ (open circles)
for varying $s_m$ in a system with 16 layers,
showing the sharp transition from coupled behavior at 
$s_m \le 1.2$ to decoupled behavior at $s_m > 1.2$.
(c) Transient interlayer correlation response $C_z$, 
for: $s_m=2.0$ and $f_d = 0.006 f_{0}^{*}$ (heavy line); 
$s_m=0.7$ and $f_d = 0.010 f_{0}^{*}$
(light line).
}
\label{fig:trans}
\end{figure}

For sufficiently strong disorder, the vortices 
in this model show a sharp 3D-2D
decoupling transition as a 
function of $s_m$
or $H$ \cite{FirstPaper22}.
A dynamic 2D-3D transition can also occur 
\cite{FirstPaper22}.
In the inset of Fig.~\ref{fig:trans}(b)
we show the critical current $f_c$
and $z$-axis correlation $C_z$ as a function of interlayer coupling $s_m$,
illustrating that a sharp transition from ordered 3D flux lines 
to disordered, decoupled 2D pancakes occurs at $s_{m}^{c}=1.2$. 
Here $f_c$ is obtained by summing $V_x = (1/N_v)\sum_{1}^{N_v}{v_x}$
and 
identifying the drive $f_d$ at which $V_x > 0.0005$, while
$C_z = 1 - \left<(|{\bf r}_{i,L} - {\bf r}_{i,L+1}|/(a_0/2)) \
\Theta(a_0/2 - |{\bf r}_{i,L} - {\bf r}_{i,L+1}|\right>$, 

\hspace{-13pt} 
where
$a_0$ is the vortex lattice constant.
The ordered phase has a much lower critical current, 
$f_{c}^{o} = 0.0008f_{0}^{*}$ than the
disordered phase, $f_{c}^{do} = 0.0105f_{0}^{*}$.  

To observe transient effects, we supercool the lattice by annealing 
the system at $s_{m} < s_{m}^{c}$ into a disordered, decoupled configuration.
Starting from this state, at $t=0$ we set $s_{m} > s_{m}^{c}$ such 
that the pancakes would be ordered and coupled at equilibrium, 
apply a fixed drive $f_d$ and observe the 
time-dependent voltage response $V_x$.
In Fig.~\ref{fig:trans}(a) we show $V_x$ for several different
drives $f_d$ for a sample with
$s_{m} = 2.0$ in a state prepared at $s_{m} = 0.5$. For 
$f_{d} < 0.0053f_{0}^{*}$ the system remains pinned 
in a decoupled disordered state.
For $f_{d} > 0.0053 f_{0}^{*}$ 
a time dependent increasing response occurs. $V_x$ does not rise
instantly but only after a specific waiting time $t_{w}$.
The 
rate of increase in $V_x$ grows as the amplitude of the
$f_{d}$ increases.  As shown in Fig.~\ref{fig:trans}(c), $C_z$ exhibits
the same behavior as $V_x$.

In Fig.~\ref{fig:trans}(b) we show a superheated system 
prepared at $s_m=2.0$ in the ordered state,
and set to $s_m = 0.7$ at $t=0$.
Here we find 
a large initial $V_x$
response that decays. With larger $f_{d}$ the decay {\it takes an increasingly
long time}. The time scale for the decay is much {\it shorter} 
than the time scale
for the increasing response in Fig.~\ref{fig:trans}(a).
In the inset of Fig.~\ref{fig:ramp}(b) 
we demonstrate the presence of a 
{\it memory} effect 
by abruptly shutting off $f_d$. The vortex 

\begin{figure}
\center{
\epsfxsize = 3.2 in
\epsfbox{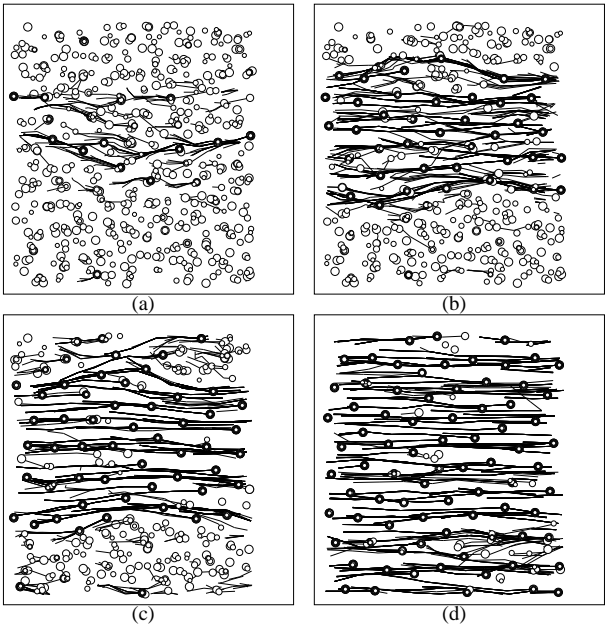}}
\caption{Pancake vortex positions (circles)  and trajectories (lines) for the 
supercooled phase
at $s_m=2.0$ and $f_d=0.007 f_{0}^{*}$, seen from the top of the sample, at
times: (a) $t=2500$; (b) $t=7500$; (c) $t=12500$; (d) $t=20000$.  
Pancakes on a given layer are represented
by circles with a fixed radius; the radii increase from the top layer
to the bottom.
An ordered channel 
forms in the sample and grows outward.}
\label{fig:cool}
\end{figure}

\hspace{-13pt}
motion stops and when $f_d$ is re-applied $V_x$ resumes at the 
same point. We find such memory on both the increasing and 
decreasing response curves. 
The response curves and memory effect seen here are very similar
to those observed in experiments
\cite{Andrei6}.

In Fig.~\ref{fig:cool} we show the vortex positions 
and trajectories in the supercooled sample with $s_m=2.0$
from 
Fig.~\ref{fig:trans}(a) for $f_{d} = 0.007 f_{0}^{*}$ 
for different times. 
In Fig.~\ref{fig:cool}(a) at $t = 2500$ the initial state is
disordered. In Fig.~\ref{fig:cool}(b) 
at $t = 7500$ significant
vortex motion occurs
through the {\it nucleation} of a single channel of moving vortices, 
which forms during the waiting time $t_w$.
Vortices outside the channel remain pinned. 
In Fig.~\ref{fig:cool}(c) at $t=12500$ the channel is wider,
and vortices inside the channel are ordered and have recoupled. 
The pinned vortices remain 
in the disordered state.
During the transient motion there is a {\it coexistence} of 
ordered and disordered states. 
If the drive is shut off the ordered domain 
is pinned but remains ordered,
and when the drive is re-applied the ordered domain 
moves again. In Fig.~\ref{fig:cool}(d) for $t = 20000$ almost 
all of the vortices
have reordered and the channel width is the size of the sample.
Thus in the supercooled case we observe
{\it nucleation} of a microscopic transport channel,
followed by {\it expansion} of the channel.  

\begin{figure}
\center{
\epsfxsize = 3.2in
\epsfbox{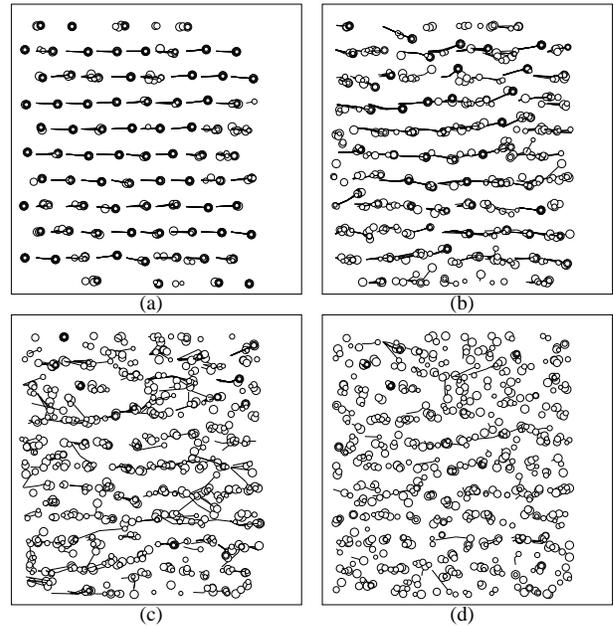}}
\caption{Vortex positions (circles)  and trajectories (lines) for the 
superheated phase at $s_m=0.7$ and $f_d=0.006 f_{0}^{*}$, at times:
(a) $t=500$; (b) $t=1500$; (c) $t=2500$; (d) $t=4500$. Each vortex line
breaks apart into pancakes as it encounters the pinning sites.
}
\label{fig:heat}
\end{figure}

\begin{figure}
\center{
\epsfxsize = 3.2in
\epsfbox{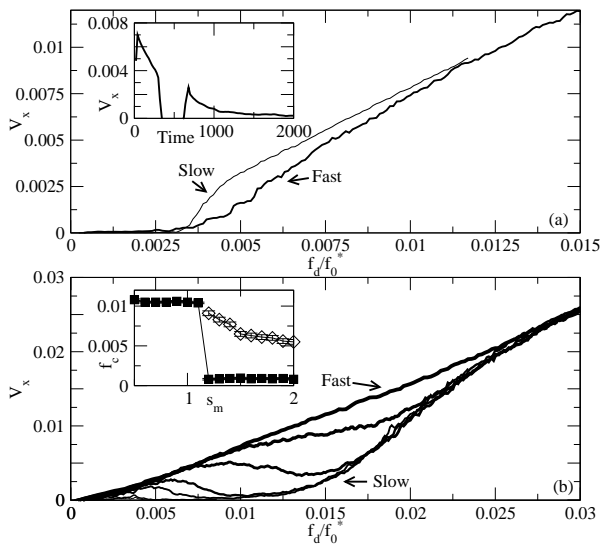}}
\caption{(a): V(I) for a supercooled system
at $s_m = 2.0$.  Light line: slow $\delta f_{d}$ of $0.0001 f_{0}^{*}$ 
every 2000 time steps. 
Heavy line: $\delta f_{d} = 0.001 f_{0}^{*}$. 
Inset to (a): Memory effect in the $V_x$ response of a superheated system 
with $s_m = 0.7$ and $f_d = 0.009f_{0}^{*}$.  The current is shut off
for 400 time steps. 
(b): V(I) for a superheated system at
$s_m = 1.1$.  From left to right, $\delta f_{d} = $
(fast) $0.02 f_{0}^{*}$, $0.01 f_{0}^{*}$, $0.005 f_{0}^{*}$,
$0.002 f_{0}^{*}$, $0.001 f_{0}^{*}$, $0.0002 f_{0}^{*}$, and 
$0.0001 f_{0}^{*}$ (slow).
Inset to (b): Effect of supercooling on $f_c$.
Filled squares: equilibrium $f_c$. 
Open diamonds: $f_c$ for samples 
prepared in a disordered, decoupled state.
}
\label{fig:ramp}
\end{figure}

The vortex positions and trajectories for a superheated sample with
$s_m = 0.7$ and $f_d = 0.006f_{0}^{*}$, as in Fig.~\ref{fig:trans}(b),
are shown in Fig.~\ref{fig:heat}(a-d).
In Fig.~\ref{fig:heat}(a) the initial vortex state is ordered. 
In Fig.~\ref{fig:heat}(b-d) the vortex lattice becomes

\hspace{-13pt}
disordered and pinned in a {\it homogeneous} manner
rather 
than through nucleation.
Each vortex line is decoupled by the point pinning
as it moves 
until the entire line dissociates and is pinned. 

We next consider the effect of changing the rate $\delta f_{d}$
at which the driving force 
is increased on $V(I)$ in both
superheated and supercooled systems.
Fig.~\ref{fig:ramp}(a) shows $V_x$ versus $f_d$, which is 
analogous to a V(I) curve, for 
the supercooled system at $s_m=2.0$ 
prepared in a disordered state.  
$V_x$ remains low during a fast ramp,
when the vortices in the strongly pinned disordered state 
cannot reorganize into the more ordered state. 
There is also considerable hysteresis 
since the vortices reorder at higher drives producing a
higher value of $V_x$ during the ramp-down.
For the slower ramp the vortices have time 
to reorganize into the weakly pinned ordered state, 
and remain ordered, producing {\it no hysteresis} in 
V(I).

In a superheated sample, the reverse behavior occurs.
Fig.~\ref{fig:ramp}(b) shows V(I) curves at different $\delta f_{d}$
for a system with $s_m = 0.7$ prepared in the ordered
state.  Here,  the fast ramp has a {\it higher} value of $V_x$ 
corresponding to the ordered state
while the slow ramp has a
low value of $V_x$.  During a slow initial ramp 
in the superheated state the 
vortices gradually disorder
through rearrangements but there is no net vortex flow through the sample.
Such a phase was proposed by Xiao {\it et al.} \cite{Xiao7}
and seen in recent experiments on BSCCO samples 
\cite{Konczykowski16}.  
At the slower $\delta f_{d}$,
we find {\it negative} $dV/dI$ characteristics 
which resemble those seen in
low- \cite{Frindt14,Borka24} and high- \cite{deGroot25} 
temperature superconductors.
Here, $V(I)$ initially increases 
as the 

\hspace{-13pt}
vortices flow in the ordered state,
but the vortices decouple as the lattice moves,
increasing $f_c$ and dropping
$V(I)$ back to zero, 
resulting in an N-shaped characteristic.

To demonstrate the effect of vortex lattice
disorder on the critical current,
in Fig.~4 we plot the equilibrium $f_c$ 
along with $f_c$ obtained
for the supercooled system, in which each sample is prepared in a state
with $s_{m} = 0.5$, and then $s_m$ is raised to a new value above 
$s_{m}^{c}$ before $f_c$ is measured.  The disorder
in the supercooled state produces a value of $f_c$ 
between the two extrema observed in the equilibrium state.
Note that the sharp transition in $f_c$
associated with equilibrium systems is now smooth.

Our simulation does not contain a surface 
barrier which can inject disorder at the edges. Such an effect
is proposed to explain 
experiments in which AC current pulses induce an increasing
response as the vortices reorder but DC pulses
produce a decaying response \cite{Andrei6,Paltiel8}. 
We observe no difference between AC and DC drives. 

In low temperature superconductors, 
a rapid increase in $z$-direction vortex wandering 
occurs simultaneously with vortex disordering \cite{Ling15}, 
suggesting that the change in $z$-axis correlations may be 
crucial in these systems as well.  Our results, along with recent 
experiments
on layered superconductors, suggest that the transient 
response seen in low temperature materials should also appear
in layered materials.

In summary we have investigated transient and metastable states near
the 3D-2D transition by supercooling or superheating the system. We find
voltage-response curves and memory effects that are very similar to those
observed in experiments, 
and we identify the microscopic vortex dynamics associated with
these transient features.  
In the supercooled case the vortex motion 
occurs through nucleation of a channel of ordered moving vortices followed
by an increase in the channel width over time.
In the superheated case the ordered phase homogeneously disorders. We also
demonstrate that the measured critical current depends on 
the vortex lattice preparation
and on the current ramp rate.  

We acknowledge helpful discussions with 
E. Andrei, S. Bhattacharya, X. Ling, Z. Xiao, and E. Zeldov. This
work was supported by CLC and CULAR (LANL/UC) by NSF-DMR-9985978, and by
the Director, Office of Adv.\ Scientific Comp.\ Res., Div.\ of Math., 
Information and Comp.\ Sciences, U.S.\ DoE contract DE-AC03-76SF00098.

\end{document}